\documentclass[%
 reprint,
 amsmath,amssymb,
 aps,
]{revtex4-2}

\usepackage{graphicx}
\usepackage{dcolumn}
\usepackage{bm}
\usepackage{comment}

\usepackage{braket}

\usepackage{color}
\usepackage{enumerate}

\begin{document}

\preprint{APS/123-QED}

\title{Electromagnetic transition amplitude for Roper resonance from holographic QCD}

\author{Daisuke Fujii}
 \email{daisuke@rcnp.osaka-u.ac.jp}
\author{Akihiro Iwanaka}
 \email{iwanaka@rcnp.osaka-u.ac.jp}
\author{Atsushi Hosaka}%
 \altaffiliation[Also at ]{Advanced Science Research Center, Japan Atomic Energy Agency, Tokai, Ibaraki 319-1195 Japan.}
  \email{hosaka@rcnp.osaka-u.ac.jp}
\affiliation{%
 Research Center for Nuclear Physics(RCNP), Osaka University, Ibaraki 567-0048, Japan.\\
}%

\date{\today}

\begin{abstract}

The Roper resonance, the first excited state of the nucleon, 
is one of the best established baryon resonances. 
Yet, its properties have not been consistently explained by 
effective models of QCD, such as the non-relativistic quark model.  
In this letter, we propose an alternative approach 
in the Sakai-Sugimoto model that is one of the holographic models of QCD. 
In particular, we analyze the helicity amplitude of the electromagnetic transitions 
at the leading of ’t~Hooft coupling $1/\lambda$.  
The model incorporates baryon structure at short distance 
by non-linear mesons surrounded by meson clouds at long distance. 
We demonstrate that the recently observed data 
by CLAS are explained  in the present approach.  

\end{abstract}

\maketitle

The Roper resonance $N^*(1440)$ is the first excited state 
of the nucleon with spin-parity $J^P=1/2^+$. 
Despite the known spin and parity quantum numbers, 
its properties have not been explained consistently by the standard model of hadrons, 
that is the non-relativistic quark model. 
Since it was first suggested by L. D. Roper in the 1960's~\cite{Roper:1964zza}, 
its mass, smaller than the negative parity state $N^*(1535)$, 
has long been a puzzle 
because the quark model predicts the mass of the Roper resonance 
larger than that of $N^*(1535)$. 

Turning to the electromagnetic interaction, 
the photo- and electro-production experiments revealed 
a further difficulty of the non-relativistic quark model~\cite{Capstick:1992xn,Capstick:1994ne}. 
In particular, the transverse helicity amplitude $A_{1/2}$ 
of the Roper resonance at the real photon point 
cannot be reproduced; 
$A_{1/2}$ value of the quark model is significantly smaller than 
the experimental ones even with the wrong sign. 
A problem is also in the strong interaction process~\cite{Arifi:2020yfp}; 
the experimentally observed large decay width of the one pion emission cannot be explained.  
Generally, transition amplitudes of various resonances 
are not easily explained by the non-relativistic quark model.

It was pointed out that relativistic effects of the confined quarks at short distance 
and pion cloud effects at long distance are 
important to improve the above mentioned problems; the mass ordering, 
the electromagnetic and strong interaction transitions
~\cite{Kubota:1976ft,Burkert:2017djo,Suzuki:2009nj,Arifi:2021orx}. 
For instance, the reason for the failure of the electromagnetic and strong transitions 
in the non-relativistic quark model 
is that the leading terms are suppressed 
due to the orthogonality of the wave functions of 
the nucleon and the Roper resonance in the long wave length limit $q \to 0$, 
where $q$ is the momentum carried either by the photon or the pion.  

In this letter, we propose an alternative but viable approach based on the holographic 
model for baryons that is the Sakai-Sugimoto model.
The model supports instantons for baryons in the four-dimensional space with an extra dimension~\cite{Douglas:1995bn,Witten:1998xy}, 
which is known to be reduced to the Skyrme model~\cite{Skyrme:1962vh}.  
In this picture, the pion cloud effect is naturally accommodated at long distance, while 
the quark dynamics at short distance by the non-linear structure of an instanton.  
In this paper it is shown that the electromagnetic transitions 
of the Roper resonance are well reproduced.  
Combining the present results with the previous successes in the mass and pion decay, 
we discuss that the holographic approach provides 
an effective method for hadrons that incorporates
important features of low energy QCD.  

The Sakai-Sugimoto model (SS model) realizes spontaneous breaking of chiral symmetry 
in terms of brane dynamics 
and has been very successful in explaining light flavor hadron physics~\cite{Sakai:2004cn,Sakai:2005yt}. 
The action in the SS model is composed of 
the Yang-Mills term $S_{YM}$ and the Chern-Simons term $S_{CS}$, 
\begin{eqnarray} \label{action}
S=S_{YM}+S_{CS}
\end{eqnarray}
where 
\begin{eqnarray}
S_{YM}=-\kappa\int d^{4}xdz{\rm tr}\left[\frac{1}{2}h(z)\mathcal{F}^{2}_{\mu\nu}+k(z)\mathcal{F}^{2}_{\mu z}\right], \notag \\
S_{CS}=\frac{N_{c}}{24\pi^{2}}\int_{M^{4}\times \mathbb{R}}\omega_{5}(\mathcal{A}), \ \ \kappa=\frac{\lambda N_{c}}{216\pi^{3}}=a\lambda N_{c}. \label{action}
\end{eqnarray}
Here, $N_{c}$ is the number of colors, $\lambda$ the 't~Hooft coupling 
and the indices $\mu,\nu=0,1,2,3$ are for the 4-dimensional space-time.
The curvatures along the extra dimension $z$ are derived from the D4-D8 brane construction to $h(z)=(1+z^{2})^{-1/3}$ and $k(z)=1+z^{2}$. 
The 1-form 
$\mathcal{A}=A_{\alpha}dx^{\alpha}+\hat{A}_{\alpha}dx^{\alpha}$ 
consists of the flavor SU(2) part $A_{\alpha}$ and the U(1) part $\hat{A}_{\alpha}$ with $\alpha=0,1,2,3,z$, 
and the field strength is 
$\mathcal{F}_{\alpha\beta}=\partial_{\alpha}\mathcal{A}_{\beta}-\partial_{\beta}\mathcal{A}_{\alpha}+i[\mathcal{A}_{\alpha},\mathcal{A}_{\beta}]$ 
with $(-,+,+,+,+)$ convention.
The Chern-Simons 5-form is given by 
$\omega_{5}(\mathcal{A})={\rm tr}\left(\mathcal{A}\mathcal{F}^{2}-i\mathcal{A}^{3}\mathcal{F}/2-\mathcal{A}^{5}/10\right)$. 
This term plays an important role in reproducing the chiral anomaly.
The hadron effective model in 4+1 dimension 
corresponds to the holographic dual of massless QCD, 
and the gauge field $\mathcal{A}_\alpha$ derived from the open string 
with both ends on the D8 brane is identified by mode expansion 
with an infinite number of vector/axial vector mesons, including pions. 

The dynamics of the baryons are dominated 
by the collective motion of the instanton~\cite{Hata:2007mb}, 
which looks quite different from the quark model 
where the baryons are described by the single particle motions of the quarks. 
Notably, in the SS model, the masses of the Roper resonances and the negative parity state 
are degenerate. 
This feature is originated from the collective motion of the baryons, 
and it is suggested that the SS model captures the features of the baryon spectrum 
better than the quark model.

Motivated by the fact,  
we have studied the decay width of one pion emission 
in the SS model and obtained encouraging results~\cite{Fujii:2021tsw}. 
This success is due to the fact that, in contrast to the quark model, 
the decay width is proportional to non-vanishing matrix element as follows, 
\begin{eqnarray}
\braket{\psi_{N^*}(\vec{x};\rho,...)|\rho^2e^{i\vec{q}\cdot\vec{x}}|\psi_N(\vec{x};\rho,...)},
\end{eqnarray}
where $\rho$ is the collective coordinate corresponding to the size of the instanton. 
The presence of $\rho^2$ evades the forbidden nature of 
the leading contribution in the non-relativistic quark model. 
This is analogous to the effect of relativistic corrections in the quark model. 

Electromagnetic excitations of nucleon resonances 
have long been studied experimentally and theoretically 
as an important source of information for understanding QCD. 
Helicity amplitudes extracted from experiments distinguish competing models 
and provide important features for understanding QCD.
In earlier years, experimental data were not sufficiently precise 
and the number of data points were not enough~\cite{Aznauryan:2004jd}. 
However in recent years, mainly with the advent of 
the Continuous Electron Beam Accelerator Facility (CEBAF) 
at the Thomas Jefferson National Accelerator Facility (JLab), 
a large amount of precise data has been obtained
~\cite{CLAS:2008roe,CLAS:2009ces,CLAS:2009tyz,CLAS:2012wxw,Mokeev:2015lda}. 
Motivated by this, several theoretical studies have also been carried out. 

The helicity amplitudes are defined by the electromagnetic current, $j^\mu_{em}$~\cite{Aznauryan:2011qj},
\begin{widetext}
\begin{eqnarray}
\label{A1/2} A_{1/2}(Q^2)=\sqrt{\frac{2\pi\alpha}{K}}\int d^3x\Braket{\psi_N,s_3=\frac{1}{2}|\epsilon^{(+)}_\mu j^\mu_{em}|\psi_ {N^*},s_3=-\frac{1}{2}}e^{i\vec{\left|k\right|}x^3}, \\
\label{S1/2} S_{1/2}(Q^2)=\sqrt{\frac{2\pi\alpha}{K}}\int d^3x\Braket{\psi_N,s_3=\frac{1}{2}|\frac{\vec{\left|k\right|}}{Q}\epsilon^{(0)}_\mu j^\mu_{em}|\psi_{N^*},s_3=\frac{1}{2}}e^{i\vec{\left|k\right|}x^3}.
\end{eqnarray}
\end{widetext}
Here, $\alpha$ is the fine structure constant,  $Q$ the four-momentum transfer of the photon, 
and the 3-momentum $\vec{k}$ of the photon is assumed to be directed along the $x^3$ axis in the $N^*$ rest frame. 
Due to the energy conservation law, we have the following equation, $k^2=Q^2+(Q^2+m^2_i-m^2_f)^2/4m^2_f$.
In the case of real photons, i.e., $Q^2=0$, we find that $\vec{|k|}$ becomes $K=(m^2_f-m^2_i)/(2m_f)$. 
The photon polarization vectors 
are defined by 
\begin{eqnarray}
&&\epsilon^{(0)}_\mu=\frac{1}{Q}(-\vec{|k|}, 0, 0, k_0) \ \ \ \ ({\rm longitudinal \ mode}) \\
&&\epsilon^{(\pm)}_\mu=\frac{1}{\sqrt{2}}(0, 1, \pm i, 0) \ \ \ \ \ \ ({\rm transverse \ mode}).
\end{eqnarray}
Therefore, to calculate this helicity amplitude, 
we prepare the wave function and electromagnetic current 
in the SS model. 

The classical solution for a four-dimensional gauge field theory 
is known as the instanton solution, 
which is usually scale-invariant. 
In the present case, the extra dimension, labeled by $z$, is curved, 
which makes the instanton solution shrink. 
However, by the repulsive nature of the U(1) term 
coupled with the Chern-Simons term, 
the classical solution of this action, the instanton solution, is stabilized. 
An analytic solution for such a system in a curved space is not known. 
However, since the size of the instanton is found to be 
proportional to $\lambda^{-1/2}$ in this model, 
the effect of this curvature is small in the large $\lambda$ limit. 
Therefore, the Belavin, Polyakov, Schwartz, and Tyupkin (BPST) instanton solution~\cite{Belavin:1975fg} 
is used as an approximate solution of the SS action~\cite{Hata:2007mb}. 
Therefore, for $M = 1, 2, 3, z$ and Pauli matrices $\boldsymbol{\tau}$.  
\color{black}
\begin{eqnarray}
&&A^{cl}_{M}\left(\mathbf{x},z\right)=-if\left(\xi\right)g\partial_{M}g^{-1} \label{A^cl_M}, \ \ A^{cl}_{0}=0, \label{SU(2)solution} \\
&&\hat{A}^{cl}_{M}=0, \ \ \hat{A}^{cl}_{0}=\frac{1}{8\pi^{2}a}\frac{1}{\xi^{2}}\left[1-\frac{\rho^{4}}{\left(\xi^{2}+\rho^{2}\right)^{2}}\right], \label{U(1)solution}
\end{eqnarray}
where $g\left(\mathbf{x},z\right)=[\left(z-Z\right)-i\left(\mathbf{x}-\mathbf{X}\right)\cdot\boldsymbol{\tau}]/\xi$, 
with $\left({\bm X}, Z \right)$ and $\rho$  the location and size of the instanton, respectively.
The profile function $f\left(\xi\right)$ is  given by $f\left(\xi\right)=\xi^{2}/\left(\xi^{2}+\rho^{2}\right)$ with $\xi=\sqrt{\left(\mathbf{x}-\mathbf{X}\right)^{2}+\left(z-Z\right)^{2}}$. 

In order to obtain the baryon wave function, 
we need to quantize the classical instanton solution~\cite{Hata:2007mb}. 
For this purpose, we first consider the motion of the instanton in the moduli space 
and make the collective coordinates time-dependent. 
Next, we consider a dynamical system 
in which the collective coordinates of the instanton are the dynamical variables, 
and perform the quantization. 
The relevant time dependent dynamical variables in the moduli space are, 
${\bm X}(t), Z(t)$, $\rho(t)$ and the  $SU(2)$  
orientation $V\left(t,x^{M};a(t)\right)$ with $V(z\rightarrow\pm\infty)\rightarrow {\bm a(t)}$ 
related to the rotational variable ${\bm a(t)}=a_{4}(t)+ia_{a}(t)\tau^{a}$ in the isospin and spin space.
We implement the time dependent collective coordinates in the gauge field as 
$A_{M}\left(t,x^{N}\right)=VA^{cl}_{M}\left(x^{N};X^{N},\rho\right)V^{-1}-iV\partial_{M}V^{-1}$. 
By substituting this gauge field for the action (\ref{action}), 
integrating over the space of $(x^{\mu},z)$, 
and quantizing these collective coordinates, 
we obtain the baryon wave functions 
\begin{eqnarray}
\psi_{N}\ &\propto& \ e^{i\vec{p}\cdot\vec{X}}\psi^{N}_{\rm radial}\left(\rho\right)e^{-\frac{M_{0}}{\sqrt{6}}Z^{2}}\left(a_{1}+ia_{2}\right), 
\notag \\
\psi_{N^{*}(1440)}\ &\propto& \ e^{i\vec{p}\cdot\vec{X}}\psi^{N^{*}}_{\rm radial}\left(\rho\right)e^{-\frac{M_{0}}{\sqrt{6}}Z^{2}}\left(a_{1}+ia_{2}\right), 
\label{wave func} 
\end{eqnarray}
where
\begin{eqnarray}
&&\label{wave_func_N} \psi^{N}_{\rm radial}\left(\rho\right)=\rho^{-1+2\sqrt{1+N^{2}_{c}/5}}e^{-\frac{M_{0}}{\sqrt{6}}\rho^{2}}, \\
&&\label{wave_func_N*} \psi^{N^{*}}_{\rm radial}\left(\rho\right)=\left(\frac{2M_{0}}{\sqrt{6}}\rho^{2}-1-2\sqrt{1+\frac{N^{2}_{c}}{5}}\right)\psi^{N}_{\rm radial}, 
\end{eqnarray}
for the spin up proton ($I_3 = 1/2, \ s_{3}=1/2$) 
with $M_{0}=8\pi^{2}\kappa$ and a finite momentum $\vec{p}$. 
Here, baryon states are labeled
by their momentum $\vec{p}$ and quantum numbers 
$(l,I_{3},s_{3},n_{\rho},n_{z})$, 
where $l/2$ is the equal isospin and spin values;
$I_{3}$, $s_{3}$ are the third components of the isospin and spin; 
and $n_{\rho}$, $n_{z}$ are the quantum numbers for oscillations along the radial and $z$-directions.  
The Roper resonance is the first radial excited state, $(l,I_{3},s_{3},n_{\rho},n_{z})=(1,1/2,1/2,1,0)$.

The electromagnetic current is defined 
as the Noether current of chiral symmetry as follows. 
Chiral transformation $(g_L, g_R)\in U(N_f)_L\times U(N_f)_R$ 
is realized by the flavor $SU(N_f)$ gauge transformation, 
$\mathcal{A}_M\rightarrow g\mathcal{A}_Mg^{-1}-ig\partial_Mg^{-1}$ with
$g(x^{\mu},z)\rightarrow g_{L/R}$ $(z\rightarrow \pm\infty)$ 
and $g(x^\mu,z)\in SU(N_f)$ and constants $(g_L, g_R)$. 

The infinitesimal local gauge transformation of this gauge symmetry 
$\delta_\xi \mathcal{A}_M(x^\mu, z)=\epsilon(x^\mu,z)\mathcal{D}_M\zeta(x^\mu,z)$, 
leads to the following five-dimensional Noether current, 
\begin{eqnarray}
&&J^M_{\zeta}=J^M_{YM\zeta}+J^M_{CS\zeta}, \\
&&J^\mu_{YM\zeta}(x,z)=-2\kappa{\rm tr}\big(h(z)\mathcal{F}^{\mu\nu}\mathcal{D}_\nu\zeta+k(z)\mathcal{F}^{\mu z}\mathcal{D}_z\zeta\big), \notag \\
&&J^z_{YM\zeta}(x,z)=-2\kappa k(z){\rm tr}\big(\mathcal{F}^{z\nu}\mathcal{D}_z\zeta\big), \notag \\
&&J^M_{CS\zeta}(x.z)=-\frac{N_c}{64\pi^2}\epsilon^{MNPQR}{\rm tr}\big(\{\mathcal{F}_{NP}, \mathcal{F}_{QR}\}\zeta\big). \notag
\end{eqnarray}
where $\zeta(x^\mu,z)$ is a $u(N_f)$ Lie algebra . 
Since the chiral symmetry of the SS model 
is related to the $SU(N_f)$ gauge transformation, 
we define the chiral current as follows~\cite{Hata:2008xc}; 
\begin{eqnarray}
j^\mu_\zeta(x)=\int^{+\infty}_{-\infty}dzJ^{\mu}_\zeta(x,z). \label{4dimcurrent}
\end{eqnarray}
Here, to satisfy the 4-dimensional current conservation law, we impose the following boundary condition, 
$J^z_\zeta(x,z\rightarrow\pm\infty)=0$. 
With $C = 0,1,2,3$ and 
\begin{eqnarray}
\psi_\pm(z)=\frac{1}{2}\pm\frac{1}{\pi}{\rm arctan}z\rightarrow
\begin{cases}
1 \ (z\rightarrow\pm\infty) \\
0 \ (z\rightarrow\mp\infty)
\end{cases},
\end{eqnarray}
we employ
$\zeta(x,z)=\psi_\pm(z)t_C$ \color{black} with $t_C=(I_2/2, \boldsymbol{\tau}/2)$, 
then the expression of the current (\ref{4dimcurrent}) leads to left/right current $j^\mu_{L/R}(x)$. 
Therefore, by substituting (\ref{SU(2)solution}) and (\ref{U(1)solution}) for (\ref{4dimcurrent}), 
the electromagnetic current 
$j^\mu_{em}=j^{\mu,C=3}_V+j^{\mu,C=0}_V/N_c$
with the vector current $j^{\mu,C}_V=j^{\mu,C}_L+j^{\mu,C}_R$ 
are given by 
\begin{widetext}
\begin{eqnarray}
\label{current_time} &&j^0_{em}(x^\mu)=\frac{3}{4\pi}\frac{\rho^2}{(r^2+\rho^2)^{5/2}}I_3+\frac{15}{16\pi}\frac{\rho^4}{(r^2+\rho^2)^{7/2}} \\
\label{current_space} j^i_{em}(x^\mu)=&&\frac{4\pi\kappa}{\rho^2}\Big(\frac{8}{r}-\frac{8r^4+20\rho^2r^2+15\rho^4}{(r^2+\rho^2)^{5/2}}\Big)\epsilon_{ijk}x^j{\rm tr}(t_k{\bm a}^{-1}t_3{\bm a})+\frac{15}{32\pi}\frac{\rho^2}{(r^2+\rho^2)^{7/2}}\big(-\epsilon_{ija}x^j\chi^a+2x^i\frac{d}{dt}{\rm ln}\rho\big)
\end{eqnarray}
\end{widetext}
where $\chi^{a}=-2i{\rm tr}\left(t^{a}{\bm a}^{-1}\dot{{\bm a}}\right)$ and $I_a$ 
is isospin operator $I_a=8\pi^2\kappa\rho^2{\rm tr}\big(i\dot{{\bm a}}{\bm a}^{-1}t_a\big)$ $(i,a=1,2,3)$.
This completes the preparation 
for calculating the helicity amplitude of the electromagnetic transition. 

There are two parameters in this model, 
Kaluza-Klein Mass $M_{KK}$ and $\kappa$, 
which we determine according to Ref. \cite{Hata:2008xc} as follows. 
For $M_{KK}$, we determine 
it to reproduce the $N$-$\Delta$ mass difference, $(1232-939)$ MeV. 
The currents used in this paper are derived 
by taking into account the leading order of $1/\lambda$. 
Therefore, we identify the nucleon mass 
with the leading term $8\pi^2\kappa$ of the mass formula. 
The $\kappa$ is then determined to reproduce the nucleon mass $939$ MeV. 
As a consequence, the two parameters are 
$M_{KK}=488$ MeV and $\kappa=0.0243$.

\begin{figure*}
\begin{minipage}{0.45\linewidth}
\centering
\hspace*{1cm}
\includegraphics[width=80mm]{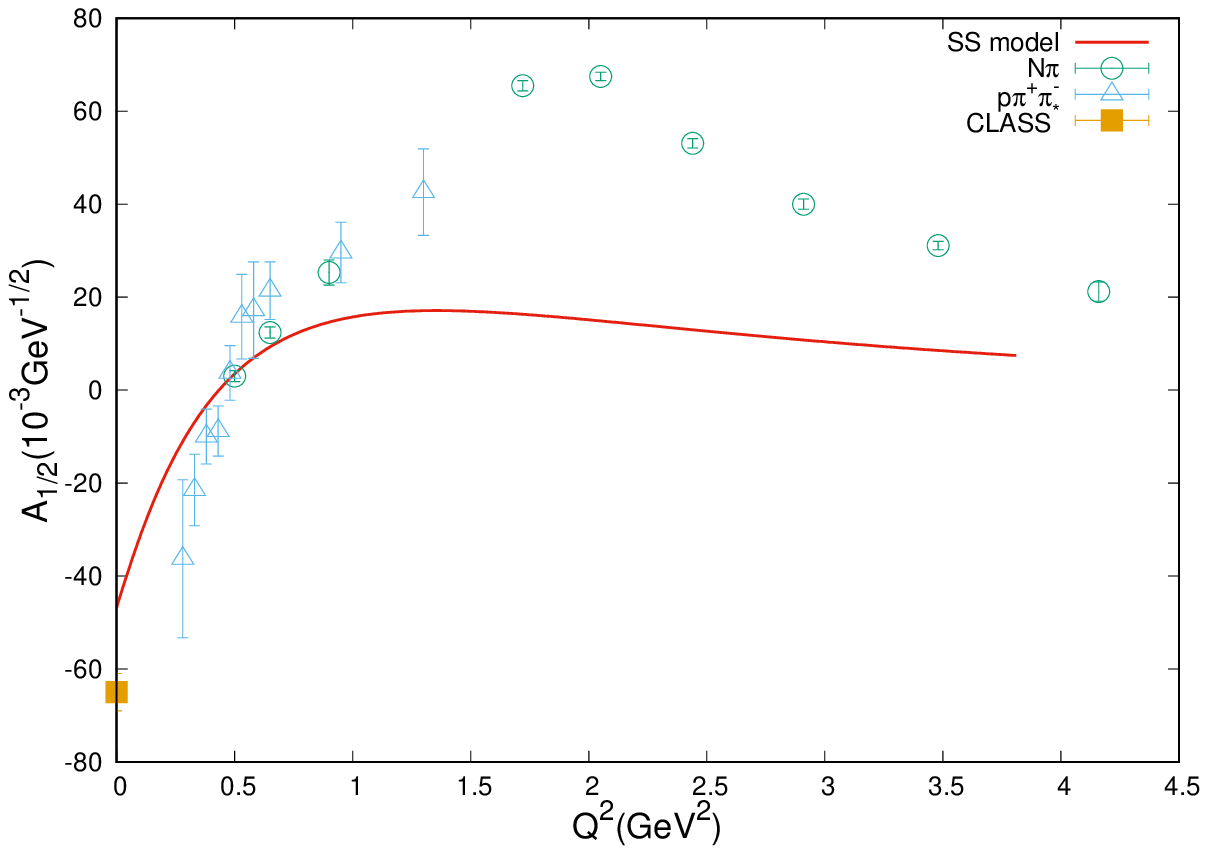}
\vskip 6mm
\caption{\label{fig_A} The transverse helicity amplitude $A_{1/2}$ in units of $10^{-3}\ {\rm GeV}^{-1/2}$ as function of 
the four-momentum transfer $Q^2$. 
The red line is result of the present study, and the sources of experimental data are shown in the panel~\cite{CLAS:2008roe,CLAS:2009ces,CLAS:2009tyz,CLAS:2012wxw,Mokeev:2015lda}.}
\end{minipage}
\hspace{0.1\columnwidth}
\begin{minipage}{0.45\linewidth}
\centering
\hspace*{1cm}
\includegraphics[width=80mm]{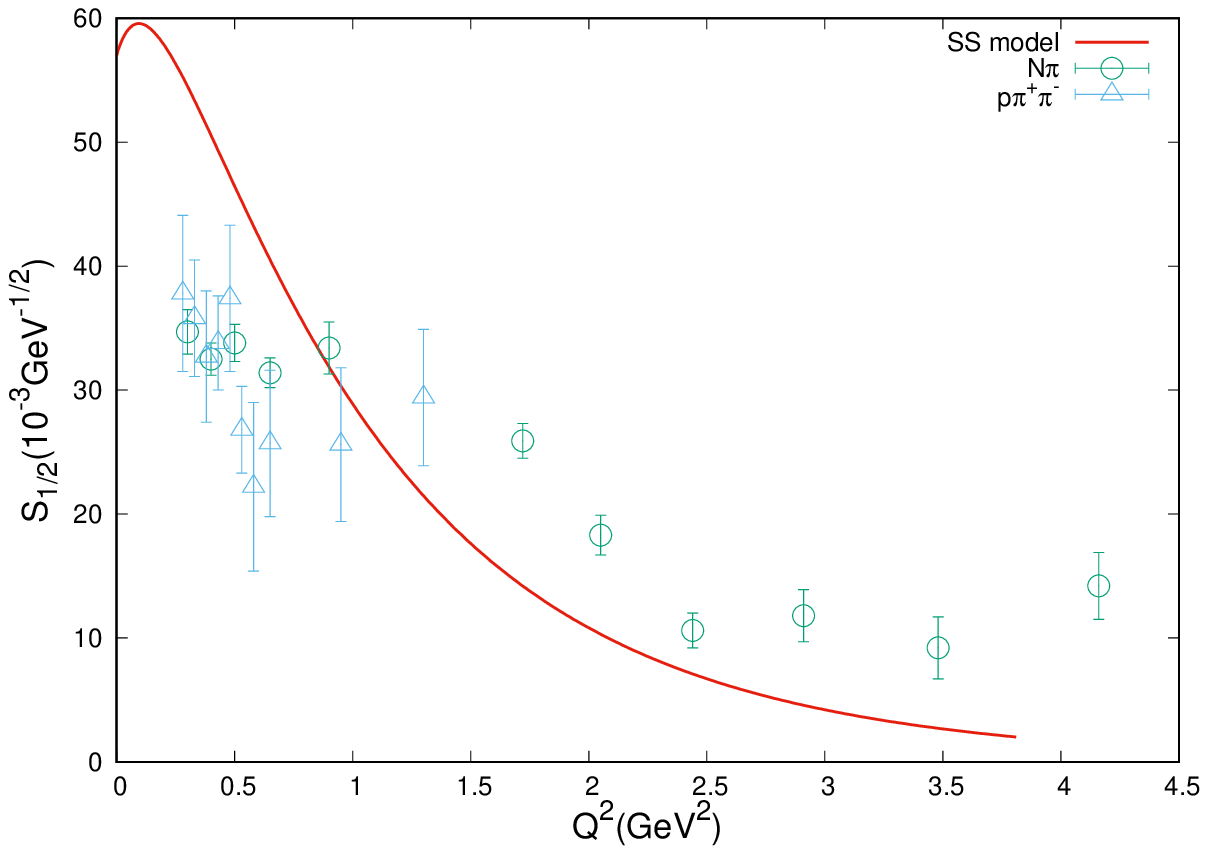}
\vskip 6mm
\caption{\label{fig_S} The longitudinal helicity amplitude $S_{1/2}$ in units of $10^{-3}\ {\rm GeV}^{-1/2}$.
The same conventions are used as in Fig.~\ref{fig_A}.}
\end{minipage}
\end{figure*}

With electromagnetic currents (\ref{current_time}) and (\ref{current_space}), 
the helicity amplitudes from the nucleon to the Roper resonance (\ref{A1/2}) and (\ref{S1/2}) 
are obtained by using the wave functions (\ref{wave_func_N}), (\ref{wave_func_N*}).
The results are shown in Figs.~\ref{fig_A} and Figs.~\ref{fig_S} as functions of $Q^2$.  
It can be seen that the helicity amplitudes obtained from our model 
with least parameters achieves global agreement with experimental data\color{black}.  
The results have some remarkable properties, as follows.

First, we note that our $A_{1/2}$ at the photon point $Q^2 = 0$ takes a finite negative value.  
The non-relativistic quark model fails to explain this property because of 
the orthogonality of the radial wave functions of the Roper resonance and the ground state nucleon.  
Several theoretical studies have been done to solve this problem, 
and it has been argued that relativistic corrections 
and the effect of the meson cloud are important~\cite{Suzuki:2009nj,Burkert:2017djo}. 
In the present approach, when the current is expanded in powers of 
$\rho$, the leading term starts at the order $\rho^2$.
The matrix element survives because of the $\rho^2$ term.  
This is the unique feature of the solitonic description of baryons where 
the collective dynamics of the meson field plays an important role.  

Second, the experimental data for $A_{1/2}$ flips its sign at around $Q^2=0.5 \ {\rm GeV}^2$. 
Our result captures this behavior very well. 
However, our model calculation underestimates 
the experimental data of $A_{1/2}$ at $Q^2 \gtrsim 1 \ {\rm GeV}^2$. 
This is because our results are calculated up to order $1/\lambda$. 
The first term in (\ref{current_space}) is the order of $\lambda$, 
and the second term is the order of $1/\lambda$. 
Moreover, the model by meson fields should be applied to the 
low energy region  $Q^2 \leq 1$ GeV$^2$ 

Third, our prediction reproduces well the experimental data for $S_{1/2}$ with sufficient strength. 
In the calculation of $S_{1/2}$, 
we can use only the time component of the current (\ref{current_time}) 
because of the current conservation law. 
We consider that the reason that our prediction of $S_{1/2}$ with sufficient strength is that 
it is dictated by the conserved charge $\int d^3xj^0_{em}=I_3+\frac{1}{2}$.  

Finally, we emphasize that there are only two parameters in this model, 
$\kappa$ and $M_{KK}$. 
We have determined these parameters from the masses of the nucleon and the delta.  
With this set of parameters, static properties of nucleons has been studied
with good agreement with experimental data~\cite{Hata:2008xc}. 
It is emphasized that by tuning only two parameters the present approach 
explains well not only static properties but also the dynamical properties of baryons.
In the present analysis, the Roper resonance has been described as 
radial density modulation.  
Hence the results provide information on the stiffness or compressibility of hadronic matter, 
which is an essential input for the study of high density matter.  

This work is supported in part by  Grants-in-Aid for Scientific Research on Innovative Areas (No. 18H05407).

\appendix

\bibliography{roper_elemag}

\end{document}